\documentclass[prl,preprint,superscriptaddress,showpacs]{revtex4}
\usepackage{graphicx}
\usepackage{epsfig}
\usepackage{amssymb}
\usepackage{mathrsfs}
\usepackage{amsmath}
\usepackage{color}
\usepackage{hyperref}

\begin{document}

\title{Confinement of Anomalous Liquids in Nanoporous Matrices}
\date{20 March 2012 --- slsfb20mar.tex}

\author{Elena~G.~Strekalova}
\affiliation{Center for Polymer Studies and Department of Physics,
  Boston University, Boston, Massachusetts 02215, USA}

\author{Jiayuan~Luo}
\affiliation{Center for Polymer Studies and Department of Physics,
  Boston University, Boston, Massachusetts 02215, USA}

\author{H.~Eugene~Stanley} 
\affiliation{Center for Polymer Studies and Department of Physics,
  Boston University, Boston, Massachusetts 02215, USA}

\author{Giancarlo~Franzese}
\affiliation{Departament de F\'{\i}sica Fonamental, 
Universitat de Barcelona, Diagonal 645, 08028 Barcelona, Spain}

\author{Sergey~V.~Buldyrev}
\affiliation{Department of Physics, 
  Yeshiva University, 500 West 185th Street, New York, New York 10033, USA}

\begin{abstract}

Using molecular dynamics simulations, we investigate the effects of
different nanoconfinements on complex liquids---e.g., colloids or
protein solutions---with density anomalies and a liquid-liquid phase
transition (LLPT).  In all the confinements, we find a strong depletion
effect with a large increase in liquid density near the confining
surface.  If the nano confinement is modeled by an ordered matrix of
nanoparticles (NPs), we find that the anomalies are preserved. On the
contrary, if the confinement is modeled by a disordered matrix of NPs,
we find a drastically different phase diagram: 
the LLPT shifts to
lower pressures and temperatures, and the
anomalies become weaker, as the disorder increases.
We find that the density heterogeneities induced by the disordered
matrix are responsible for the weakening of the LLPT and the
disappearance of the anomalies.
 

\end{abstract}

\maketitle

Many experiments in recent years have shown that a number of liquids
exhibit highly anomalous properties \cite{vilaseca2011}. The data for
liquid metals, metalloids, nonmetals, oxides and alloys---including Ga,
Bi Te, S, Be, Mg, Ca, Sr, Ba, SiO$_2$, P, Se, Ce, Cs, Rb, Co, Ge,
Ge$_{15}$Te$_{85}$---colloids, protein solutions, organophosphates, such
as triphenyl phosphite (TPP), AY20 melts [(Al-O)$_{80}$--(Y-O)$_{20}$]
and water, reveal the presence of a temperature of maximum density (TMD)
below which the density decreases under isobaric cooling
\cite{vilaseca2011}.  In a number of these systems, such as P, TPP, and
AY20 \cite{P}, it has been shown the existence of a liquid-liquid
phase transition (LLPT)  ending in a liquid-liquid critical point
(LLCP) between two coexisting liquids with the same composition but
different structure: the high density
liquid (HDL) and the low density liquid (LDL).  Data from experiments on
silica, C, Se, Co, and water are consistent with a LLPT \cite{sio2}.
Here we ask how the structure of the nanoconfinement may change the
anomalous behavior of the liquid and affect the LLPT and the
LLCP. This question is relevant 
across a wide range of nanotechnological applications, biological
systems, and is of general interest for phase transitions in confined
systems \cite{referee}.

We model the liquid using two different potentials, (i) the Jagla ramp
potential \cite{Jagla} and (ii) the continuous shoulder potential
\cite{Fr}, which reproduce thermodynamic and dynamic anomalies, LLPT and LLCP
in bulk.  We model the nanoconfinement by a fixed matrix of NPs
connected by bonds which the liquid particles can penetrate. 
Potential (i) has a hard-core at distance $r
= a$, and a linear ramp for $a<r\leq b$ decreasing from interaction
energy $U_R>0$ to $-U_0<0$, plus a linear ramp for $b<r\leq c$ increasing
from $-U_0$ to 0.  We adopt $b/a=1.72$, $c/a=3$ and $U_R=3.56
U_0$. The liquid particles interact with NPs via hard core repulsion at distance
$r_0\equiv (a+D_{\rm NP})/2$, where $D_{\rm NP}$ is the NP diameter.
Potential (ii) has a repulsive shoulder and an attractive well with
energy minimum $U_0$, with parameters chosen to fit a potential proposed
in Ref.~\cite{HG93}.  The interaction with NPs is given by a
$1/(r-r_0)^{100}$ power law.

For both potentials we perform simulations at constant number $N$ of
liquid particles, constant volume $V$, and constant temperature $T$,
with periodic boundary conditions.  For (i) we employ a discrete
molecular dynamics (MD) algorithm by discretizing the linear ramp
potential into steps, with $\Delta U\equiv U_0/8$ \cite{Buldyrev}.  For
(ii) we use a standard MD with a velocity Verlet integrator and the
Allen thermostat \cite{Fr}.

\begin{figure}
{\includegraphics[scale=0.15]{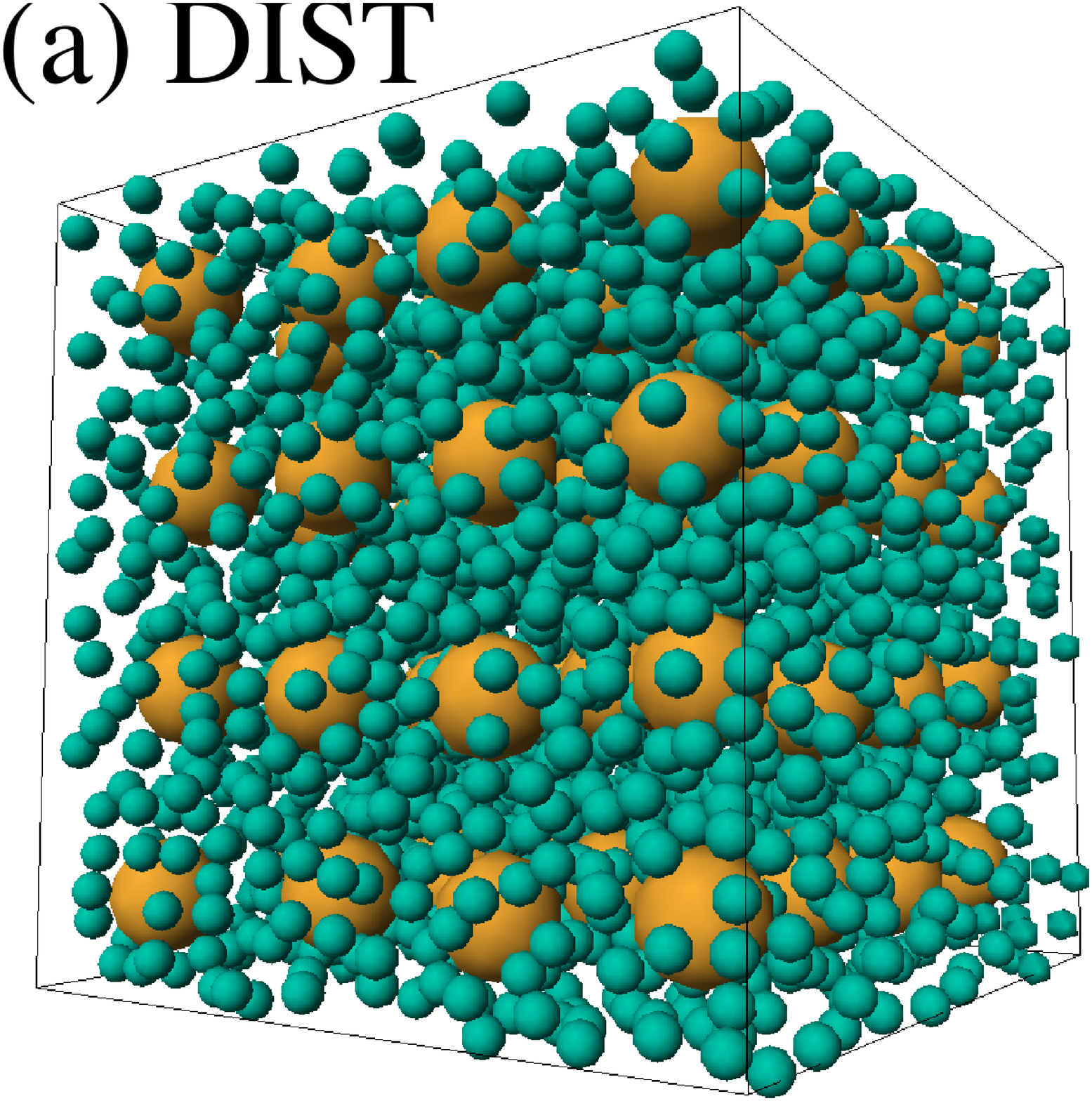}
 \hspace{0.4in}
 \includegraphics[scale=0.15]{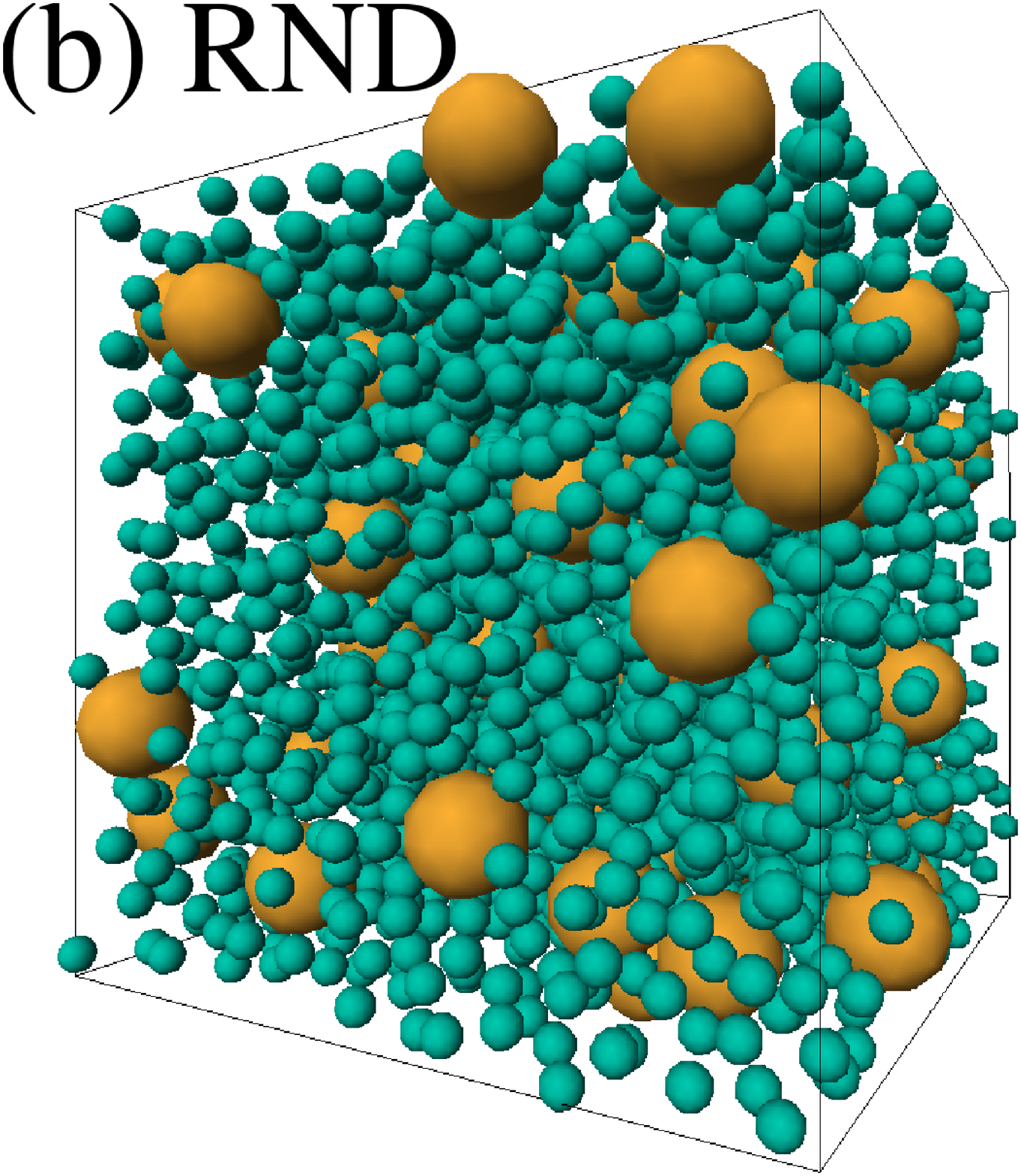}}
\includegraphics[scale=0.4]{fig1cd.eps}
\caption{Effect of confinement.  Snapshots of the anomalous liquid
  (green) confined in a fixed matrix of NPs (yellow) in a DIST (a) and
  RND (b) configuration.  Polynomial fits of simulated isochores of
  densities $0.89 \leq \rho/\rho_c^{\rm bulk} \leq 1.59$ (bottom to top
  in the one-phase region) for DIST (c) and RND (d).  Randomness reduces
  the temperature and pressure of the LLCP (circles), the separation
  between the HDL (lower filled triangles) and LDL spinodals (upper open
  triangles) and the separation between the TMD (diamonds) and the
  temperature of minimum densities (TminD, squares). Samples of error
  bars on $P$ are given in panel (c).  Lines connecting symbols are
  guides for the eyes.
\label{P-T}}
\end{figure}

We consider three different structures for the matrix: a perfect cubic
lattice (CUBE); a cubic lattice with Gaussian distortions (DIST) with a
standard deviation equal to 1/4th the separation between centers of NPs,
which still preserves an approximately periodic and ordered structure of
the confinement (Fig.~\ref{P-T}a); and a completely random (RND)
configuration of NPs obtained by simulating a gas of hard spheres
(Fig.~\ref{P-T}b).  The volume fraction of NPs is $x_{\rm NP}\equiv
V_{\rm NP}/V$, where $V$ is the volume of the cubic simulation box and
$V_{\rm NP}=N_{\rm NP} 4\pi r_0^{3}/3$ is the volume inaccessible to the
liquid.  Our results here, if not otherwise indicated, are for liquid
(i) confined by the matrix of $N_{\rm NP}=64$ NPs with diameter $D_{\rm
  NP}/a=3$ at $x_{\rm NP}=24.5\%$ and $V/a^3=20.6^3$.  We control the
density $\rho\equiv N/(V-V_{\rm NP})$ of the liquid particles by
changing $N$ in the interval between 1845 and 3887. We take into account
that the excluded volume rescales the pressure $P$ by $V/(V-V_{\rm
  NP})$.  We find that the results for liquid (ii) are consistent in
similar conditions.

\begin{figure}
\includegraphics[scale=0.4]{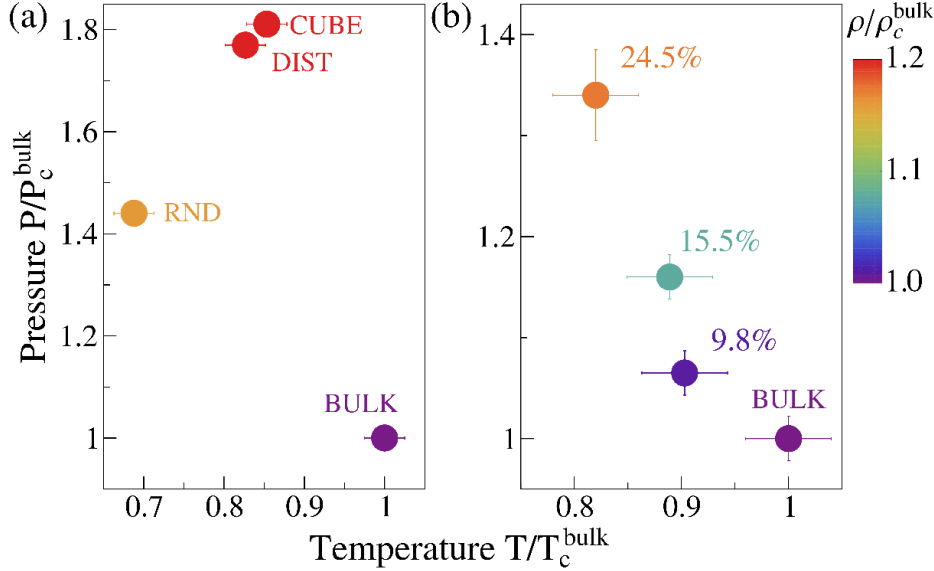}
\caption{The effect of different confinements on the parameters of the
  LLCP.  Color-coded circles represent the LLCP parameters in the
  $P-T-\rho$ phase space (a) for the liquid confined in the fixed matrix
  of NPs with CUBE, DIST and RND configuration.  Increasing disorder in
  the confinement, from CUBE to DIST to RND, shifts the LLCP down in
  $\rho$, $T$, and $P$.  (b) Upon decreasing concentration $x_{\rm NP}$
  (label near the symbols) for the CUBE confinement of the liquid (ii),
  the LLCP approaches the bulk case.  Here we use $N_{\rm NP}=64$ NPs
  with $D_{\rm NP}/a=3$ in $V/a^3=20.6^3$ at $x_{\rm NP}=24.5\%$ with
  $1452\leq N\leq 2508$ (with spontaneous crystallization below the
  LLCP); or in $V/a^3=24^3$ at $x_{\rm NP}=15.5\%$ with $2570\leq N\leq
  4439$; or in $V/a^3=28^3$ at $x_{\rm NP}=9.8\%$ with $4358\leq N\leq
  7528$.  We find the same behavior for liquid (i).
\label{cr.pt}}
\end{figure}

For liquid (i), the bulk system displays a LLCP at $k_BT_c^{\rm
  bulk}/U_0=0.375$, $P_c^{\rm bulk}a^3/U_0=0.243$, and $\rho_c^{\rm
  bulk} a^3=0.37$ \cite{Jagla}.  Figures~\ref{P-T}(c) and \ref{P-T}(d)
show simulated isochores for DIST and RND confinement, respectively,
with the HDL-LDL spinodal lines calculated using conditions $(\partial P
/\partial \rho )_T =0$ and $(\partial ^2 P /\partial \rho ^2 )_T \neq
0$, and the LLCP obtained at the point of merging of the spinodal lines
where $(\partial P /\partial \rho )_T = (\partial ^2 P /\partial \rho
^2)_T=0$. We find that every confinement causes the LLCP to shift to a
lower $T$, a higher $\rho$, and a higher $P$ than in the bulk liquid
(Fig.~\ref{cr.pt}a).  As the disorder in the confining matrix increases,
the $T$ shift is more pronounced and the $\rho$ and $P$ shifts less
pronounced.  We find the same qualitative trend in the LLCP shifts for
liquid (ii), and that the LLCP progressively approaches the bulk case
when the NP concentration decreases (Fig.~\ref{cr.pt}b), consistent with
previous results for NP-liquid mixtures \cite{CorradiniPRE10}.

While the periodic DIST confinement preserves the LDL-HDL coexistence
region observed in bulk liquid (Fig.~\ref{P-T}c), which is consistent
with a strong first-order LLPT, the RND confinement shrinks the
coexistence region (Fig.~\ref{P-T}d) and weakens at the LLPT, which
manifests itself in the shrinking of the region between the spinodals in
the $P$--$T$ plane. This shrinking is qualitatively consistent with that
found for a model of water in a random hydrophobic pore-like confinement
\cite{Strekalova}.

The region of density anomaly is bounded by the lines of the TMD and the
temperature of minimum density (TminD) located by the extrema of the
isochores. In the bulk system the TminD line for high densities is
hindered by the glass temperature line and cannot be observed in the
equilibrium liquid.  Here we observe that the periodic structure of the
confinement can dramatically affect density anomaly manifestations.
Compared to the bulk, confinement decreases TMD and increases TminD,
shrinking the $T$ range of the density anomaly. The density anomaly is
still well defined in the DIST case, but it appears much less pronounced
in the RND case. For a RND matrix of $N_{\rm NP}=19$ large confining NPs
with diameter $D_{\rm NP}/a=5$ at $x_{\rm NP}=24.5\%$ and
$V/a^3=20.6^3$, the TMD and TminD are completely absent (not shown).

\begin{figure}
\includegraphics[scale=0.4]{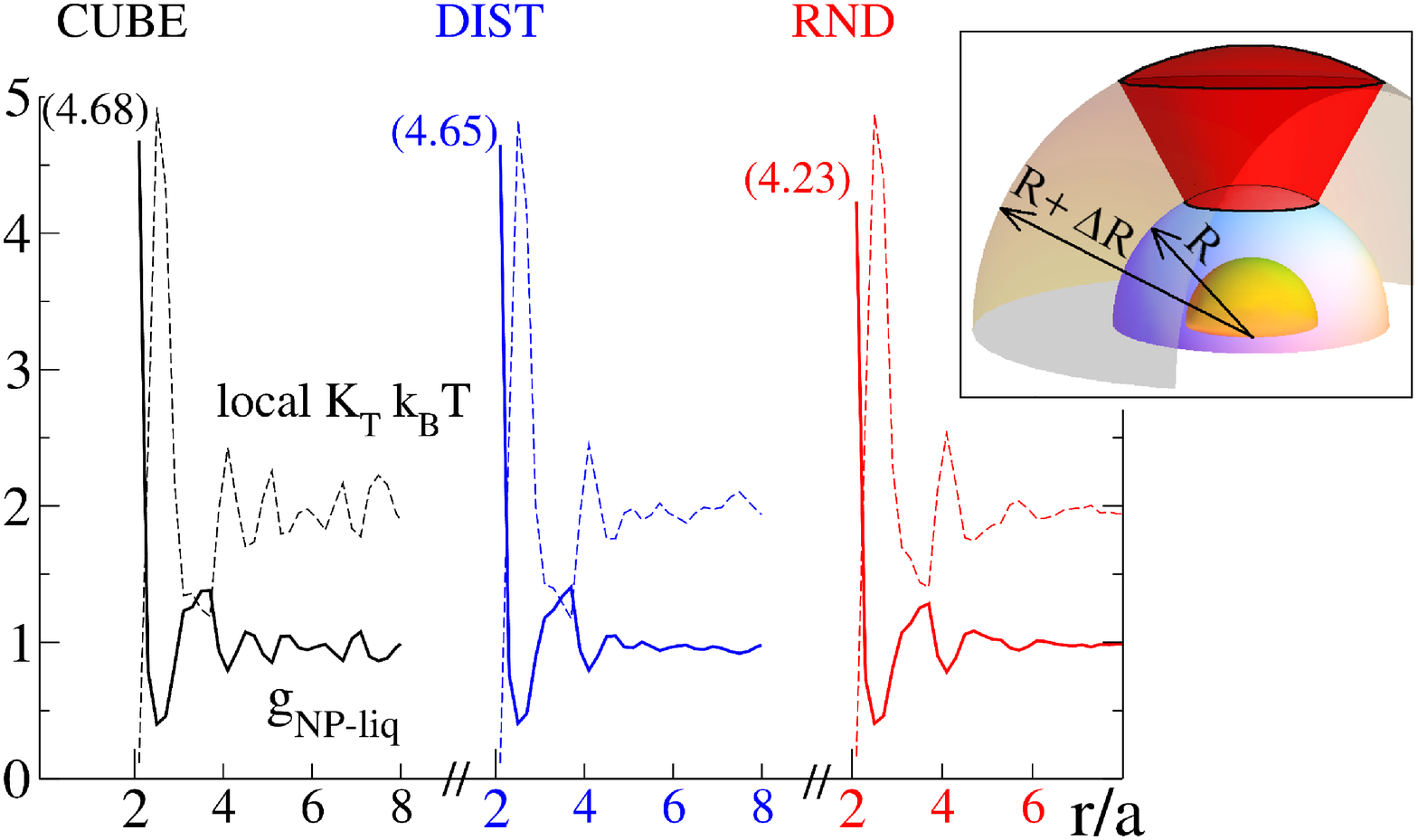}
\caption{The liquid adsorbs onto the NPs.  The fluid density profile 
  $g_{\rm NP-liq}(r)$ at $T/T_c^{\rm bulk}=1.12$
  for CUBE (leftmost), DIST (center) and RND (rightmost) confinements
  for density $\rho/\rho_c^{\rm bulk} = 1.59$ (solid lines) display
  large maxima (values in parenthesis) at the closest NP-liquid particle
  distance $r=r_0\equiv 2a$.  Local compressibility $K_T$ (dashed lines)
  show large peaks near the minimum of $g_{\rm NP-liq}(r)$.  The results
  for different confinements are shifted horizontally for clarity.
  Inset: Schematic representation of calculation of $g_{\rm NP-liq}(r)$
  and local $K_T$ inside equal-volume ($\Delta W=2.77a^3$)
  conical regions between two concentric spheres with different radii
  $R$ and $R+\Delta R$ centered at the NP (yellow), where $\Delta R=0.2a$, and
$R=2.0a,2.2a,...8.0a$ (one such segment is shown in red). The axis of
  the segment is chosen at random $10,000$ times for each NP.  $g_{\rm
    NP-liq}(r)$ is computed by counting the number of liquid particles
  and local $K_T = (\frac {\langle n^2  \rangle}{\langle n \rangle ^2}
  -1) \frac{\Delta W}{k_BT}$ from fluctuations of number of liquid
  particles $n$. 
\label{gr-N}}
\end{figure}

To understand the origin of the different effects of the different
confinements, we study the density of the liquid in the vicinity of
NPs. We find that a layer of liquid adsorbs onto the NPs, as revealed by
the fluid density profile $g_{\rm NP-liq}(r)$(Fig.~\ref{gr-N}).  We
understand the increase of density near the NP surface as a
consequence of entropy maximization.  By packing near the fixed NPs,
the adsorbed liquid particles allow more free space to the the rest of
the liquid, maximizing the entropy of the system (depletion effect).
This result evokes a similar effect found for water at confining 
surfaces, regardless of the hydrophobic or hydrophilic interaction with 
the surface \cite{Garde}, and for hard-sphere fluids in contact with purely 
repulsive particles \cite{Konig}, showing that the increase of contact 
density is not related to specific interactions or anomalous behaviors 
and making a bridge between water and simple fluids.

We find that, by increasing randomness in the confinement, the
probability of overlap of NP exclusion volumes increases and the
depletion effect decreases. As a consequence, the density of liquid near
the NPs decreases (Fig.~\ref{gr-N}).  In addition, we analyze the
density fluctuations and the associated measurable response function,
the local isothermal compressibility $K_T$ (Fig.~\ref{gr-N}), of the
liquid in the vicinity of the NPs.  We find that $K_T$ is extremely
small at the interface, consistent with a tight packing of liquid
particles around the NPs. Near the first minimum of $g_{\rm NP-liq}(r)$,
$K_T$ is, instead, twice as high as in the bulk.  A high local density
causes the density increase of the LLCP (Fig.~\ref{cr.pt}) because, when
part of the liquid is adsorbed onto the NPs, an average liquid density
larger than bulk is necessary to build up the critical fluctuations.
The shift is more pronounced for CUBE and DIST confinement, with respect
to RND, because the more ordered the confinement the larger the NP
surface available for the depletion effect.

\begin{figure}
\includegraphics[scale=0.4]{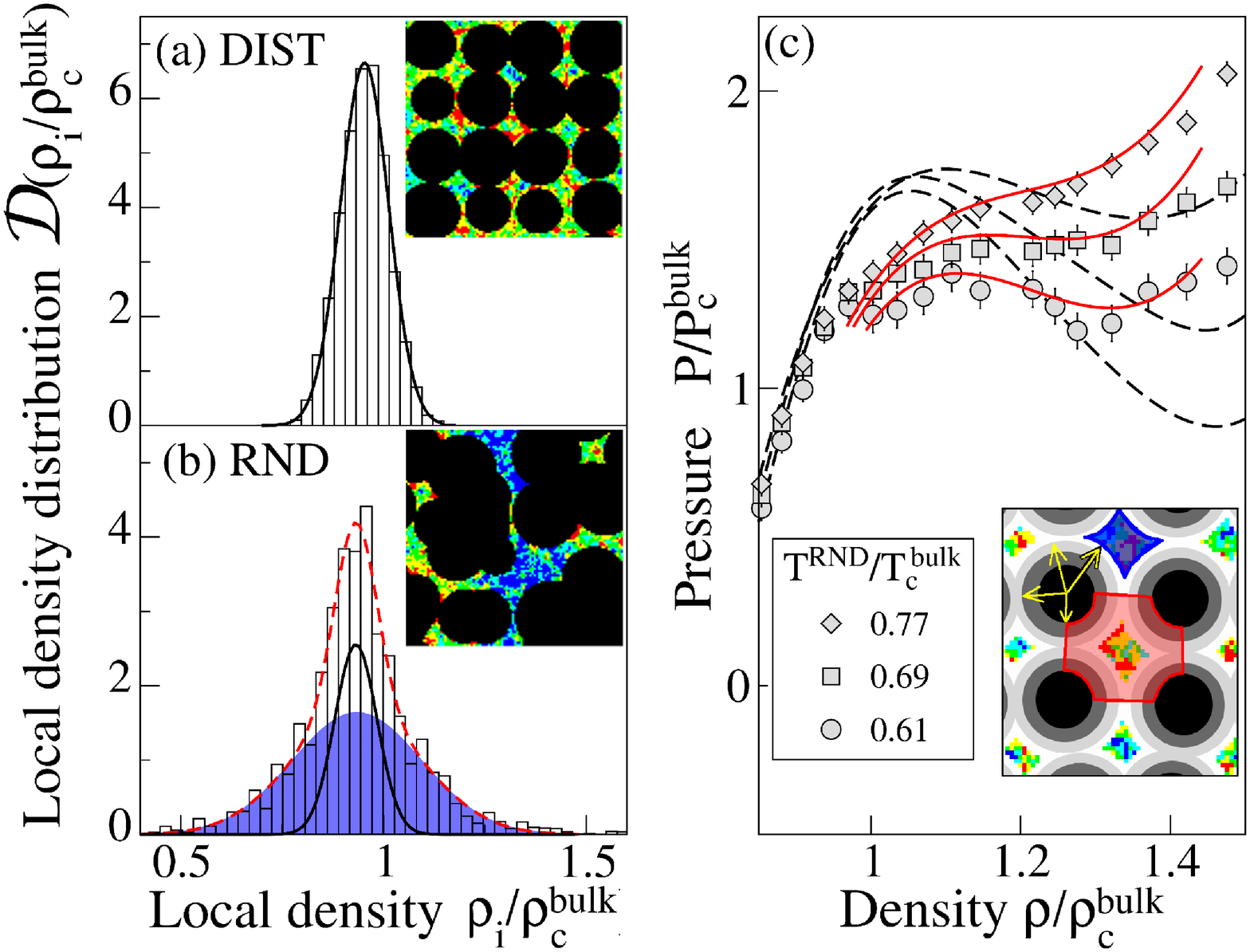}
\caption{The distribution of local density $\mathcal{D}(
  \rho_i/\rho_c^{\rm bulk})$ of the liquid inside the pockets for global
  liquid density $\rho/\rho_c^{\rm bulk}=0.94$ and $T/T_c^{\rm
    bulk}=0.88$. (a) In the DIST confinement at $P/P_c^{\rm bulk}=1.3$,
  $\mathcal{D}^{\rm DIST}(\rho_i/\rho_c^{\rm bulk})$ is a Gaussian
  centered $\rho/\rho_c^{\rm bulk}=0.94$ and with standard deviation
  $\sigma_{D}=0.055$.  Inset: cut through the simulation box. The liquid
  density (hight to low color-coded from blue to red) is computed inside
  spheres of radius $1.5a$ that do not intersect NPs.  Areas for which
  we can not evaluate liquid density with this method are in black.  (b)
  In RND confinement, at $P/P_c^{\rm bulk}=1.25$, the broad
  $\mathcal{D}^{\rm RND}(\rho_i/\rho_c^{\rm bulk})$ (red dashed line) is
  the result of two Gaussian components, both centered in
  $\rho/\rho_c^{\rm bulk}=0.94$, but with different standard deviations:
  one is due to the local density fluctuations (black line) with
  $\sigma_{R1}=0.052$, as in DIST, and the other with
  $\sigma_{R2}=0.159$ (shaded) due to the heterogeneity in pockets
  volumes. Inset: as in panel (a), but for RND.  (c) Calculation of
  $P^{\rm RND}$ by taking into account the component of
  $\mathcal{D}^{\rm RND}(\rho_i/\rho_c^{\rm bulk})$ due to the
  heterogeneity in pocket volumes.  Polynomial fits of the isotherms,
  $P^{\rm DIST}(\rho)$ at constant $T$ (black dashed lines: from top to
  bottom $T/T_c^{\rm bulk}=0.77$, 0.69, 0.61), are used in Eq.~(\ref{1})
  to get an estimate of $P^{\rm RND}(\rho)$ at the same $T$ (red solid
  lines), which compare well, within a range of densities close to the
  LLPT, with the simulation data for RND (symbols).  Inset: 2D
  representation of the exclusion spheres (black, grey, white circles)
  with their radii $r_e$ (yellow arrows), which for $r_e=r_0$ (black
  circles) coincide with NPs. A cavity is highlighted in blue diamond
  and a pocket in red. For clarity, the liquid (colored regions) is
  shown inside cavities only.
\label{rho-v}}
\end{figure}

To better understand how confinement structure affects the physical
properties of a liquid, we study the liquid's local density distribution
inside the confinement matrix.  We identify the region not occupied by
the NPs and partition it into disconnected {\em cavities\/} (inset
Fig.~\ref{rho-v}c) based on the Delaunay tessellation algorithm
described in Ref.~\cite{Sastry}. We define the exclusion spheres
concentric with NPs and gradually increase their radius $r_e$ with a
small step $\Delta r_{\rm e}=0.1a$. We designate the space not occupied
by exclusion spheres as {\it void\/} of size $r_{\rm e}$ and denote it
$\Omega(r_{\rm e})$. For $r_{\rm e}=r_0$, $\Omega(r_0)$ is a connected
set for both RND and DIST confinements. The volume of $\Omega(r_0)$ is
equal to $V-V_{\rm NP}$.

In DIST confinement, when $r_{\rm e}>4.1a$, $\Omega(r_{\rm e})$ breaks
into 64 small disconnected cavities, associated with 64 distorted cubic
{\it pockets\/} formed by 8 adjacent NPs.  The volume $\omega_i$ of each
pocket $i=1, \dots, 64$ is given by the volume of all Delaunay
tetrahedra comprising the corresponding pocket minus the volume occupied
by the NPs forming the pocket. We define the particle density of liquid in
each pocket $\rho_i\equiv N_i/\omega_i$, where $N_i$ is the number of
liquid particles inside pocket $i$. We find that the volumes $\omega_i$
are narrowly distributed, with the local liquid density distribution
$\mathcal{D}^{\rm DIST}(\rho_i)$ given by a Gaussian with variance
$\sigma^2_{D}$ (Fig.~\ref{rho-v}a).

In RND confinement, $\Omega(r_{\rm e})$ remains fully connected up to
$r_{\rm e}=4.2a$. As we increase $r_{\rm e}$, small pockets break away
from the largest part of $\Omega(r_{\rm e})$ one by one.  When $r_{\rm
  e}=5.4a$ we count, for different random configurations, approximately
60 pockets, for which we calculate $\omega_i$ and $\rho_i$, finding a
large variety of sizes and shapes.  We compute $\mathcal{D}^{\rm
  RND}(\rho_i)$ and find that in RND it can be approximated with the sum of two
Gaussian distributions: one similar to the DIST case with $\sigma_{R1}
\approx \sigma_{D}$ and the other resulting from the heterogeneity of
volumes $\omega_i$ of the pockets with $\sigma_{R2} > \sigma_{R1}$
(Fig.~\ref{rho-v}b).

We hypothesize that in RND confinement the observed pressure $P^{\rm
  RND}(T,\rho)$ results from averaging local pressures in each
pocket. At temperature $T$ we estimate $P^{\rm RND}$ using the average
of the $P^{\rm DIST}(T,\rho_i)$ over all heterogeneous pockets
(Fig.~\ref{rho-v}c),
\begin{equation}
P^{\rm RND} (T,\rho)= \int 
P^{\rm  DIST}(T, \rho+\rho\xi) \frac{\exp[-\xi ^2/2\sigma_{R2} ^2]}{\sqrt{2 \pi
    \sigma_{R2}}} d \xi.
\label{1}
\end{equation}
Due to averaging over different densities $\rho_i\equiv \rho+\rho\xi$,
the non-monotonic subcritical isotherm $P^{\rm DIST}(\rho)$ at
$T=T_c^{\rm RND}<T_c^{\rm DIST}$ becomes a monotonic critical isotherm
$P^{\rm RND}(\rho)$ that closely fits the simulation results for the RND
confinement in the vicinity of the LLPT. Thus our averaging technique
allows us to reproduce quantitatively the differences we found when we
compared DIST and RND confinements, i.e., the critical temperature,
pressure, and density decrease (Fig.~\ref{cr.pt}a) and density anomaly
region shrinks (Fig.~\ref{P-T}c,d).  Thus the presence of density
heterogeneity and the reduced depletion effect in the RND confinement
matrix give us the key to understanding the effect of confinement
structures. It is important to stress the differences of the effect of
confinement on the LLPT and the liquid-gas phase transition (LGPT).
While in both cases the critical temperature is significantly reduced,
the effects of random confinement and ordered confinement are practically
indistinguishable in the case of LGPT. This is because in LGPT, the density of
liquid particles has a much smaller increase near NPs than in LLPT. Thus in LGPT
randomness does not lead to local density heterogeneities, which
produce a strong effect on the LLPT.

In conclusion, we predict that anomalous liquids with a LLPT retain
their bulk phase diagram and density anomalies when they are confined in
a porous matrix with an ordered structure.  Furthermore, when there is a
small distortion of the confinement, the glass temperature is reduced
with respect to bulk, allowing the direct observation of the TminD
locus. A strong depletion effect induces a large increase of density in
the vicinity of the NPs. The effect is smaller when the confinement has
a random structure. Randomness induces heterogeneity in the local
density, which weakens the LLPT, narrows the LLPT coexistence region, and
washes out the density anomalies.

Although the anomalous liquids considered here are in principle 
different from water, our results could qualitatively explain recent 
experiments for confined water, the prototypical anomalous liquid. 
While the TminD
locus has been observed in supercooled water under hydrophilic
confinement by the MCM-41 silica nanoporous matrix
\cite{Mallamace-density-min}, its absence has been reported in the
hydrophobic mesoporous material CMK \cite{Zhang09}.  MCM-41 forms a
regular matrix \cite{Mallamace-density-min}, but CMK consists of grains,
each with a disordered pore structure \cite{Zhang09}.  This suggests
that the disparity of results for different confinements may arise from
the different amount of disorder in the confining structures,
independent of the interaction details of the anomalous liquid.

\bigskip

\noindent We thank D.~Corradini, P.~Gallo, S.~Sastry and K.~Stokely for
discussions.  EGS, JL, and HES acknowledge the support of NSF grants
CHE0908218 and CHE0911389, SVB the Dr. Bernard W. Gamson Computational
Science Center at Yeshiva College, and GF the MICINN grant FIS2009-10210
(co-financed FEDER) and the EU FP7 grant NMP4-SL-2011-266737.

\end{document}